# DNA unzipped under a constant force exhibits multiple metastable intermediates

Claudia Danilowicz*, Vincent W. Coljee†, Cedric Bouzigues*, David K. Lubensky‡, David R. Nelson*, and Mara Prentiss*§

*Department of Physics, Harvard University, Cambridge, MA 02138; †Symphogen A/S, DK-2800 Lyngby, Denmark; and ‡Bell Labs, Lucent Technologies, Murray Hill, NJ 07974

Single molecule studies, at constant force, of the separation of double-stranded DNA into two separated single strands may provide information relevant to the dynamics of DNA replication. At constant applied force, theory predicts that the unzipped length as a function of time is characterized by jumps during which the strands separate rapidly, followed by long pauses where the number of separated base pairs remains constant. Here, we report previously uncharacterized observations of this striking behavior carried out on a number of identical single molecules simultaneously. When several single λ phage molecules are subject to the same applied force, the pause positions are reproducible in each. This reproducibility shows that the positions and durations of the pauses in unzipping provide a sequence-dependent molecular fingerprint. For small forces, the DNA remains in a partially unzipped state for at least several hours. For larger forces, the separation is still characterized by jumps and pauses, but the double-stranded DNA will completely unzip in less than 30 min.

**T**he separation of double-stranded DNA (dsDNA) into single-stranded DNA (ssDNA) is fundamental to DNA replication in living organisms, and, of course, to the PCR. In equilibrium, DNA will separate when the free energy of the separated ssDNA is less than that of the dsDNA. In most studies of DNA separation, the strands are separated by increasing the temperature of the sample until the DNA melts. In living organisms, however, DNA separation is not thermally driven. Rather, enzymes and other proteins force the two strands apart. Recent work has begun to investigate the *force*-induced separation of dsDNA at temperatures where dsDNA is thermally stable in the absence of an applied force.

One might naively predict that, if one unzips DNA by applying a constant force above the thermodynamic unzipping transition, the dsDNA would fairly rapidly separate into ssDNA regardless of the sequence; however, DNA sequence variations can produce metastable intermediate states corresponding to an incomplete unzipping of the DNA. When the dsDNA is unzipped at a constant force, these metastable states result in significant pauses during which the number of separated base pairs does not change with time despite the constant applied force. The lifetime of these metastable states can be much longer than an hour. In this paper, we report the observation of such pauses in the unzipping of dsDNA by a constant force.

Most DNA melting experiments measure only the average dynamical behavior, losing information about intermediate states and individual variation. Recent single molecule experiments (1–10) have studied the dynamics of the unwinding, stretching, or unzipping of strands of DNA by an applied force, shedding light on movements, interactions, configuration changes, and other dynamical properties of individual molecules that may differ significantly from the average (11). The sequence dependence of the unzipping of DNA by an applied force may provide information about the origins of replication along DNA molecules. The sequence specificity of the replication origins, such as A + T rich areas, has been generally recognized as an important feature, especially in prokaryotes and some simple eukaryotes (12). Replication also involves major topological changes in DNA. In particular, in bacteria, the progression of the replication fork requires two topoisomerases (gyrase and topoisomerase IV) to relieve the mechanical strain imposed by the unwinding of the parental DNA duplex, therefore resolving DNA knots and entanglements. Single molecule experiments have also focused on the role of topoisomerases (8, 13).

DNA unzipping experiments have usually been performed either at constant displacement (9, 10) or at varying loading rates (14). When the *displacement* is held constant (9, 10), the force adjusts to compensate for the different average binding energies in AT-rich and GC-rich regions. Hence, one does not expect large jumps and metastable states in this case. An approach that is more easily modeled theoretically and may be more closely related to strand separation in cells is one where constant *force* is applied to separate the strands. For homopolymeric DNA, the unzipping transition is expected to occur continuously and completely at a constant rate once the constant applied force exceeds the threshold for separating the single base pairs. The behavior of coding dsDNA with a heterogeneous sequence, however, will be closer to that of a random heteropolymer than of a homopolymer. A long heteropolymer unzipped by a constant applied force will not unzip continuously at a constant rate, but will instead unzip discontinuously, pausing at a series of energy minima where the strand separation will cease until an energy barrier is overcome. Therefore, the number of base pairs opened in a dsDNA is expected to show sharp jumps, as a function of time, that depend on the applied force as well as the base sequence (15, 16). But even for identical DNA molecules, the number of base pairs that separate at a given time will vary because the separation requires random thermal activation that occurs differently in different identical molecules.

The DNA unzipping problem has been addressed in several theoretical publications (15–20). Refs. 15 and 16 provide a detailed analysis of *heteropolymer* DNA unzipping in a constant force ensemble and describe important differences with constant extension experiments. According to this description, the unzipping process will exhibit a series of long plateaus followed by large jumps, thus showing several "microphase transitions" where DNA partially unzips until it encounters an energy barrier and the process pauses. For a random sequence, these barriers scale roughly as $\Delta\sqrt{m}$, where $\Delta$ is a typical GC/AT energy variation ($\approx k_B T$) and $m$ is the number of base pairs. Related phenomena have been observed in simulations of the mechanical denaturation of proteins (17).

The unzipping can continue if thermal fluctuations overcome the barriers or if the force is increased; if the applied force is high enough, it is possible to overcome all of the barriers easily and the DNA will unzip in a short time. The energy landscape at a lower force will be strongly sequence dependent, with different locations of the energy minima for different random sequences. A variety of semimicroscopic models have been used to describe DNA unzipping without, however, considering sequence heter-ogeneity (18–21). Although interesting dynamical effects and a barrier to initiation of unzipping have been uncovered, the integrated effects of sequence randomness in long DNA strands can produce barriers that are hundreds of times larger than those considered in these papers (15, 16). Although most of the theoretical work has emphasized 1D models, the complex 3D topology of the DNA may also contribute to the dynamics of unzipping at constant force. In this work, we report observations of the phase transition between dsDNA and ssDNA from phage λ induced by applying a constant force to separate the two strands.

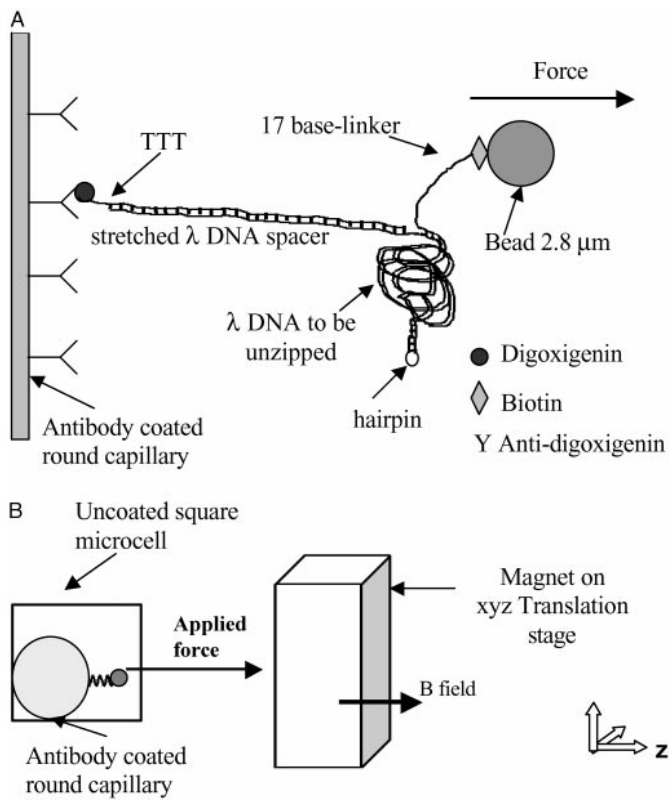

**Fig. 1.** Molecular construction and square cell. (*A*) Schematic of the DNA binding to the inner glass capillary and the magnetic bead such that pulling the bead away from the surface will cause the dsDNA shown on the right side of the diagram to be separated into two single DNA strands. Note that the figure is not to scale, considering that λ DNA contains 48,502 bp. (*B*) Schematic of the side view of the square capillary containing the round glass capillary to which the DNA molecules are bound. The magnetic tweezer apparatus exerts the controlled force on the magnetic beads, a microscope is used for observation, and two thermoelectric coolers are used to control the temperature of the sample during the initial incubation. The magnetic beads are pulled to the right in a direction parallel to the bottom and top surfaces of the square capillary, and perpendicular to the surface of the round capillary at a height equal to the radius of the round capillary, where we focus the microscope. This design allows us to view DNA molecules that are offset from the surfaces of the square capillary, and to infer the number of separated base pairs by measuring the separation between the magnetic bead and the surface of the round capillary.

## Methods

**Sample Preparation.** A molecular construction similar to the one reported in ref. 9 was prepared, as shown in Fig. 1*A*. The construction presents two λ DNAs (New England Biolabs) as covalently connected molecules, one behaving as a long linker or spacer arm and the second as the DNA to be opened. The DNA to be opened had a hairpin added to one end. The linker DNA was hybridized and ligated with a digoxigenin-labeled oligonu-cleotide at one end while the DNA to be opened was hybridized and ligated with a biotinylated oligonucleotide at one end and with an oligonucleotide forming a hairpin at the other end to prevent the separation of the two strands when complete unzipping was achieved. Finally both DNA molecules were hybridized and ligated to obtain the two covalently connected DNA molecules. The two differences between the construction used in this work and that reported in ref. 9 are as follows: additional three single-stranded T nucleotides between the end of the dsDNA spacer and the digoxigenin label that binds the spacer to the glass, and a 42-bp hairpin that is longer than the 22-bp hairpin reported in ref. 9, were used. Comparison of the construction without the three single-stranded T nucleotides as described in ref. 9 with one including these nucleotides showed that our construction increased the number of molecules tethered to the surface from 4% to 45%. This significant increase in adhesion is likely due to the strong reduction in steric hindrance that is encountered in the interaction of antibody to antigen or streptavidin to biotin in the presence of bulk materials.

The DNA construction connected a superparamagnetic bead to a glass surface using specific interactions as schematically represented in Fig. 1*A*. The DNA sample was contained in a glass microcell with a square cross section of 0.6 mm, connected to polypropylene tubing for sample injection. A round glass capillary, 0.33 mm in diameter, and closed at its ends, was placed in the square microcell, defining a straight line and allowing precise measurements of the distance of the beads to the surface. The position of this inner capillary was checked with an inverted microscope, and the sample was injected when the capillary touched the bottom of the square cell (see Fig. 1*B*).

**Preparation of Modified Surfaces.** The surfaces were prepared by adsorption of molecules that can recognize target labels attached to the ends of the DNA construction. The two different labels used were digoxigenin and biotin, and the corresponding proteins were deposited on the glass and bead surfaces. The binding to the inner circular glass capillary was accomplished by using an IgG antibody specific to digoxigenin. Adsorption of antidigoxigenin IgG on glass was carried out by incubating 1 μM antibody solution in PBS (pH 7.4) overnight at room temperature. The modified surface was then successively washed with water and 33 mM Tris acetate, 66 mM potassium acetate, 10 mM magnesium acetate, 0.5 mM DTT buffer (pH 6.8). Finally, the DNA sample was mixed with a suspension of streptavidin-coated beads in buffer and incubated in the glass surface at 40°C for 15 min. Streptavidin-coated beads were obtained from commercial sources, and the choice of the bead sizes depended on the range of forces required in each experiment.

To gain insight into the stability of the binding of the DNA to the surfaces of the glass slide and the beads, we exerted force on a single dsDNA molecule tethered between the glass and a bead. The beads remained tethered at forces as high as 30 pN at room temperature. Even at a force of 30 pN, only at temperatures close to the antibody denaturing temperatures of roughly 65°C (22) was it possible to detach the beads from the surface toward the opposite side of the capillary, proving the stability of the specific bonds. Thus, the observed detachment of beads tethered to the surface by partially unzipped DNA molecules is probably due to breaks in the ssDNA, rather than to a detachment of the DNA from the surface of the glass or the bead.

**Apparatus.** In our apparatus, the magnetic field gradient is produced by one stack of five permanent magnets each of 6.4 × 6.4 × 2.5 mm³ dimension. The total magnetic field is approximately that of a solenoid with its long axis in the *z* direction, as shown in Fig. 1*B*. The field along the *z* axis is purely in the *z* direction and uniform relative to the solenoid axis to within a few percent. The magnetic force on each superparamagnetic bead was given by $m \triangledown B$, where $B$ is the magnetic field and $m$ is the magnetic moment on the bead (23). This resulting force on a given bead in the sample is almost exclusively in the z direction, and varies by <1% over the region of the liquid sample monitored in the experiment as explained in ref. 23. The magnets were held in a lateral position with respect to the microcell on a three-axis translation stage to exert a force perpendicular to the glass surface to which the DNA was bound. The magnitude of the force applied on the beads was determined by the distance between the magnet and the glass surface. The values quoted for the force are the average values for a given field gradient. For a given field gradient, differences in the iron content of the beads produce a variation of about ±20% in the force exerted on DNA molecules that are attached to different beads in the sample.

Using this experimental setup, dozens of single molecule measurements on identical DNA molecules can be accomplished in parallel, thus allowing comparisons and direct observation of sample-to-sample fluctuations. Reproducibility of pauses and jumps from (genetically identical) molecule to molecule allows us to distinguish sequence effects from transient entanglements of unwound DNA with, e.g., bits of randomly located broken DNA, protein, or dust.

Bead tracking was performed with an inverted microscope (objective lens ×10, 0.25 numerical aperture) and therefore, the position of the beads with respect to the capillary surface could be measured. It was feasible to follow the position of the superparamagnetic beads as a function of time by acquiring the images with a video frame grabber as can be seen in Fig. 2. The rounded surface of the inner capillary allowed us to optimize the focusing of the beads tethered to a line along the edge of the round capillary that was closest to the magnet. Focusing the microscope on this line allowed us to measure only those beads that were tethered along this edge, providing a well defined bead surface distance and a large separation between the dsDNA being unzipped and all surfaces of the square capillary.

## Results

The trajectory of the beads connected to the open end of the DNA construction was measured for multiple molecules at several forces. When none of the base pairs separated, the distance between the bead and the surface was 16.5 μm, corresponding to the contour length of the DNA linker that connects the strand to the surface. As the DNA unzipped, the distance between the bead and the capillary increased to a maximum length of ≈80 μm. At the beginning of the experiments when we applied a very low <1 pN force, we found that some of the beads were at a distance >16.5 μm. These beads were not followed during the experiments.

The separation between the magnetic beads and the surface of the glass inner round capillary to which they were bound was measured as a function of time. Fig. 3 shows curves corresponding to several different individual molecules. The curves with solid and dashed lines correspond to beads under 15 and 20 pN of force, respectively. In either case, there are long periods of time during which the distance of the bead from the surface (number of base pairs unzipped) is nearly constant. These regions are separated by intervals where the distance between the bead and the surface changes rapidly. We interpret the rapid change as the result of a thermal excitation sufficient to allow the force to overcome a local barrier to unzipping. Rapid unzipping then takes place in a region of the DNA where the force is sufficient to overcome the binding between the pairs of strands until a new barrier is encountered. In such regions, rezipping may also occur, so the number of unzipped base pairs can fluctuate between two nearly degenerate partially zipped configurations, separated by a barrier, as shown in Fig. 3 at 15 pN for $t >$ 2,500 s. Experiments that used short strands of RNA have observed similar bi-stability (24), but for such short strands, the only relevant minima in the energy landscape for unzipping occurred when the duplex region was either fully paired or fully separated. In the case of the much longer λ DNA, only a part of the sequence was oscillating between the zipped and unzipped states.

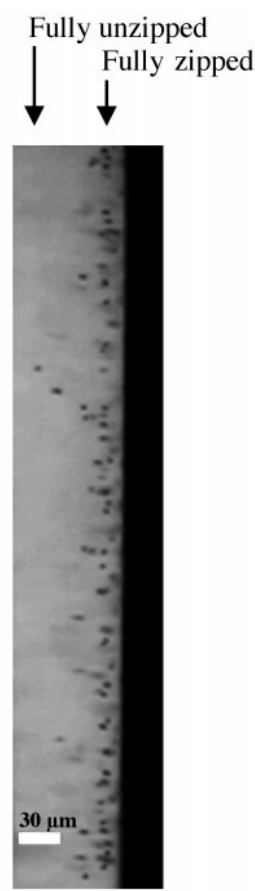

**Fig. 2.** Photograph of the top view of the square cell shown in the schematic in Fig. 1B. The dark object shown at the right is the surface of the round capillary. The dark dots are the magnetic beads tethered to the surface by DNA. The image was taken after applying a constant force of 15 pN during 45 min. The arrows at the top of the figure show the fully zipped and fully unzipped lengths of the molecules.

The intrinsic barriers $|\Delta E| \sim \Delta \sqrt{m}$, due to sequence heterogeneity for unzipping, say, 20-kb pairs of DNA when the force is at its critical value are of order 10–100 $k_B T$, with $k_B T \approx 300$ K (16). The time scale for surmounting a typical barrier is $\tau \approx \tau_0 \exp [|\Delta E|/k_B T]$, where $\tau_0$ is a microscopic time for base pair opening and closing. With $\tau_0 \approx 10^{-7}$ s (25), we see that barriers of order 20–30 $k_B T$ should be relatively easy to surmount on our experimental time scale ≈4,000 s. Higher barriers can be overcome by a slight increase of the force past the unzipping threshold.

Because the unzipping events are thermally activated, different molecules have very different numbers of separated base pairs as a function of time, although the regions where unzipping stops recur in multiple single molecules. The locations of the deep minima that cause pauses depend only on the sequence and applied force, not on the history of a particular molecule. Although the length of the pauses may differ, the pause site remains the same. This effect can be seen in Fig. 3, where each colored line represents position vs. time for a different DNA molecule. For example, the solid yellow, dark green, and light green lines represent the unzipping of three distinct individual dsDNA molecules. All three of these different individual mol-ecules of the dsDNA being unzipped at 15 pN pause at $\approx19.5$ $\mu m$ and then oscillate in length between 19.5 and 20.5 $\mu m$ for as long as several minutes. This reproducibility from molecule to molecule is even more striking given that the variation in the magnetic moment of the beads results in a distribution in the applied force of approximately $\pm20\%$ with respect to the average. This result is consistent with the theoretical prediction that the location of the deepest minimum in a given region of the energy landscape should remain the same for a range of applied forces (16). The dashed lines in the figure represent dsDNA being unzipped at 20 pN, where again different colors represent different molecules. Notice that at 20 pN the average size of the jumps where the strands easily unzip becomes longer as predicted, (16) because the larger force can more easily overcome unzipping barriers between the strands. The data taken at 20 pN is clearly distinct from that taken at 15 pN; however, even at 20 pN, none of the strands fully unzipped during the 3,600-s time range of this experiment.

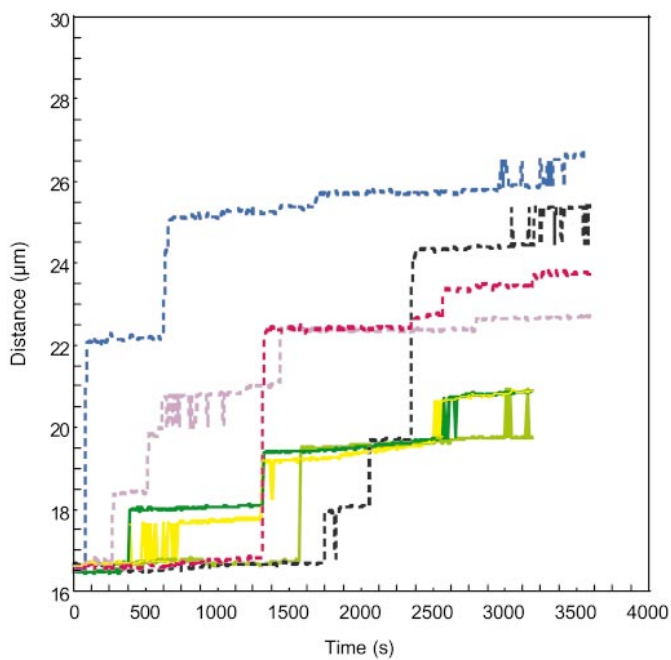

**Fig. 3.** Measured distance between the centers of the magnetic beads tethered to the surface of the glass capillary by different identical DNA molecules as a function of time. The solid lines represent trajectories for beads with 15 pN of applied force, and the dashed lines represent trajectories for beads with 20 pN of applied force.

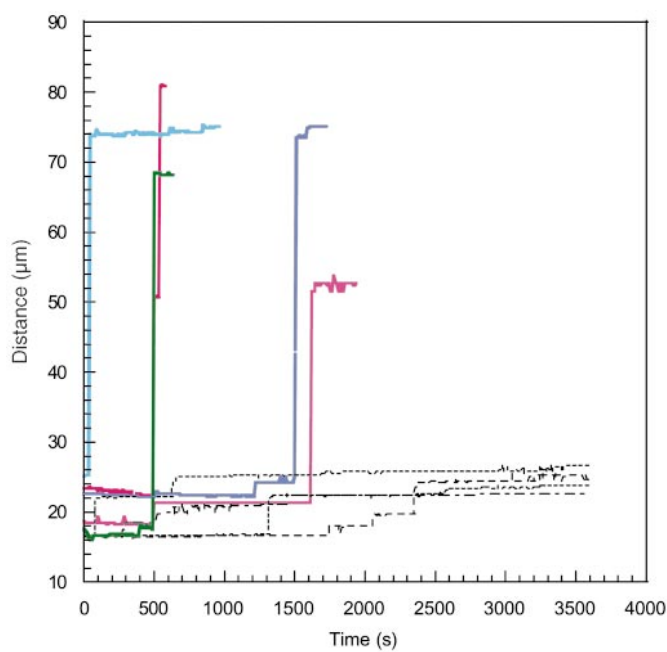

**Fig. 4.** Measured distance between the centers of the magnetic beads tethered to the surface of the glass capillary by different identical DNA molecules as a function of time. The colored solid lines represent trajectories for beads under 30 pN of applied force. The dashed lines show the 20-pN data from the previous figure on the same scale.

On the other hand, if even larger forces were applied, full unzipping does occur as shown in Fig. 4 for several molecules. The colored solid lines in the figures correspond to unzipping at a constant force of 30 pN, and the dashed lines correspond to the data shown in Fig. 3 for unzipping at a constant force of 20 pN. At the beginning of the experiment, the positions of the beads exhibited some partial unzipping that occurred before the data acquisition began, but pauses in the unzipping were nevertheless observed. Some of these pauses may be due to transient entanglements or twist relaxation instead of sequence effects. The pauses were succeeded by rapid unzipping over a much larger number of base pairs than was characteristic at 15 or 20 pN. One of the beads detached from the surface before the DNA had fully unzipped, possibly due to nicks in the backbone.

From these figures, it is clear that the unzipping of a given molecule at constant force does not proceed at a constant rate. At room temperature, for low forces ($<20$ pN), the unzipping is very discontinuous, showing long times at which there is almost no unzipping, followed by regions of rapid unzipping that then terminate in another region of very slow unzipping. In the constant force ensemble, there may be a large variation in the number of unzipped base pairs for each molecule at a given time, even when all of the molecules in the sample were subject to the same force. This behavior was predicted by theory and is understood by considering the random sequence of $\lambda$ phage DNA closer to that of a heteropolymer. In contrast, a homopolymer of DNA exposed to a constant unzipping force would unzip at a constant average rate, and the molecules in the sample would have only a small, diffusive dispersion in the number of unzipped base pairs at a given time.

A theoretical model was used to simulate the energy landscape describing DNA unzipping at constant force. These coarse-grained landscapes are shown in the left column of Fig. 5 for constant applied forces of 14.5, 15.0, and 16.3 pN. To compute the landscapes, the pairing/stacking energies for each base pair were taken from the thermodynamic data predicted for DNA oligonucleotides with a nearest-neighbor model (26). We then adapted the coarse-graining procedure of Le Doussal et al. (27, 28) to systematically eliminate small barriers until only a few large ones remained. Notice that only at 14.5 pN (corresponding to the critical unzipping force $F = F_c$ in thermodynamic equilibrium) is there an obvious peak in the averaged potential landscape, due to a high density of strong GC bonds in the first half of the sequence and a low density in the second half. Because the coarse-graining procedure was adapted to our experimental conditions of biased unzipping from left to right in the figures, there are few minima on the steep downhill GC-depleted half of the DNA. The macroscopic barrier to unzipping is extremely large, $>3,000$ kT, and there are many deep local minima that can be tens to approximately hundreds of kT deep. These local minima are sufficient to pause the unzipping even on a downhill slope, and allow the DNA to move by ratcheting through sequential local minima. Fig. 5 *Right* shows an expanded scale, which reveals coarse-grained minima in detail for applied forces of 14.5, 15.0, and 16.3 pN, corresponding to the range of base pairs selected in Fig. 5 *Left*. These minima represent the number of unzipped base pairs at which the unzipping is expected to pause. If the barrier is overcome by a thermal fluctuation, the DNA should rapidly unzip until it pauses again at another local potential barrier. At 14.5 pN, the minima are approximately evenly distributed along the first half of the sequence, with no stable states in the second half, except for the state of completely unzipped DNA. At larger applied forces, there are no obvious macroscopic barriers, although there again are a series of small local minima.

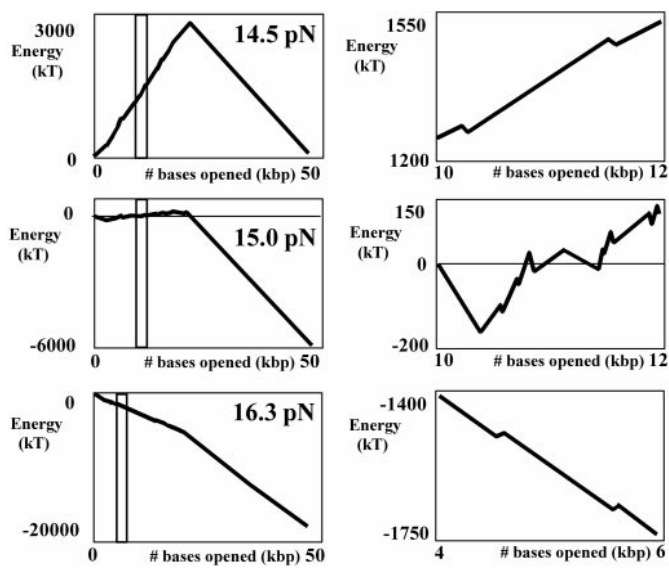

**Fig. 5.** Coarse-grained potential in kT vs. number of unzipped kilobase pairs. (*Left*) The calculated average energy landscape for DNA under a constant force: 14.5, 15.0, and 16.3 pN. (*Right*) An expanded view of the range selected in *Left*, revealing the coarse-grained minima in detail.

## Discussion

We have unzipped dsDNA using constant force and found that the unzipping is discontinuous as a function of time, exhibiting a series of jumps and plateaus, just as DNA separation *in vivo* by enzymes can occur discontinuously with long pauses at particular points in the sequence. For the simple *in vitro* case of our experiment, previous theoretical work based on a 1D model predicted our observed discontinuous unzipping and the jumps due to thermal fluctuations over local sequence-dependent energy barriers to strand separation, where the locations of the pauses are insensitive to the applied force when the applied force is close to its critical value. For forces below 17 pN, experimental results are in reasonable agreement with theory. At forces above 20 pN, theory predicts that there should be no pauses in the unzipping, although pauses are observed at forces as large as 30 pN. The disagreement between the theoretical predictions and the experimental results may be attributed to one or more of the following causes: the dispersion of applied forces due to variations in the magnetization of the beads; the complex 3D topology of coiled dsDNA (e.g., transient entanglements); contamination of the sample; and inaccuracies in the theoretical extrapolation to room temperature of the base pairing energies measured at temperatures above 50°C. Further work on the temperature and salt concentration dependence of the unzipping may provide additional insight about the mechanisms governing discontinuous force-induced unzipping of DNA *in vitro* and *in vivo*.

## Appendix

**Numerical Methods.** The plots in Fig. 5 were produced by using a method inspired by the real-space renormalization group approach to diffusion in a 1D random medium of Fisher, Le Doussal, and Monthus as described in refs. 27 and 28. The basic idea is to successively merge the smallest energy barriers into their larger neighbors, leading to a more and more coarse-grained energy landscape in which only the largest (and thus most dynamically important) barriers remain. Specifically, we began by assigning a pairing-free energy to each base pair using the parameter values for the standard 10-parameter nearest-neighbor model inferred by SantaLucia, Allawi, and Seneviratne described in ref. 26. Likewise, we calculated the free energy per base pair as a function of applied force for the unzipped ssDNA, treating the ssDNA as a freely jointed chain with Kuhn length 1.5 nm (1). An initial energy landscape $E(m)$ has been computed. $E(m)$ is defined as the difference in free energy between the fully paired λ dsDNA and the same molecule with $m$ base pairs unzipped; it is calculated by subtracting $2m$ times the free energy of an unzipped nucleotide subject to an applied force from the pairing free energy of the first $m$ base pairs. The energy landscape can be thought of as a succession of upwards or downwards steps, one for each base. We next replaced this landscape by one in which runs of neighboring upwards steps were replaced by a single energy barrier of height equal to the sum of the upwards steps (and likewise for downwards steps). Because the experiments showed primarily unzipping, with relatively little backwards motion, we modified the original decimation procedure of Le Doussal and colleagues (27, 28) to focus on upwards barriers to unzipping. We iteratively removed the smallest upward barrier and replaced it by an average of its neighbors. If the magnitude of the barrier to be decimated was larger than the magnitude of either of the adjacent downwards barriers, we grouped the sequence of up, down, up barriers containing the smaller of the two downwards neighbors into one upwards barrier with height equal to the sum of the three barriers being merged. Otherwise, the sequence of down, up, down barriers containing the upward barrier to be decimated was similarly merged into a downward barrier. An advantage of this approach compared with the original work of Le Doussal and colleagues (27, 28) is that it allows one to more efficiently pick out the largest barriers to forwards motion; disadvantages are that it sometimes leaves a small downward barrier in between two large upwards barriers undecimated and that it provides little information about the size of barriers to backward motion.

We are grateful to Professor Michael E. Fisher and, especially, Professor Daniel Branton for comments on the manuscript. We also acknowledge the contributions of Fabiano Assi and Robert Jenks in building the apparatus used in these experiments and Abhijit Sarkar for helpful suggestions. This research was supported by National Science Foundation Grants PHY-9876929 and DMR 9809363, Department of the Navy Grant N00014-01-1-0782, and Harvard University. Work by D.R.N. was also supported by National Science Foundation Grant DMR9714725.

**BIOPHYSICS**